# In-situ Characterization of Crystallization and Melting of Soft, Thermoresponsive Microgels by Small-Angle X-ray Scattering

Dmitry Lapkin,[a] Nastasia Mukharamova,[a] Dameli Assalauova,[a] Svetlana Dubinina,[a,b] Jens Stellhorn,[a,c] Fabian Westermeier,[a] Sergey Lazarev,[a,d] Michael Sprung,[a] Matthias Karg,[e] Ivan A. Vartanyants[a,f†] and Janne-Mieke Meijer[g‡]

Depending on the volume fraction and interparticle interactions, colloidal suspensions can form different phases, ranging from fluids, crystals, and glasses to gels. For soft microgels that are made from thermoresponsive polymers, the volume fraction can be tuned by temperature, making them excellent systems to experimentally study phase transitions in dense colloidal suspensions. However, investigations of phase transitions at high particle concentration and across the volume phase transition temperature in particular, are challenging due to the deformability and possibility for interpenetration between microgels. Here, we investigate the dense phases of composite core-shell microgels that have a small gold core and a thermoresponsive microgel shell. Employing Ultra Small Angle X-ray Scattering, we make use of the strong scattering signal from the gold cores with respect to the almost negligible signal from the shells. By changing the temperature we study the freezing and melting transitions of the system *in-situ*. Using Bragg peak analysis and the Williamson-Hall method, we characterize the phase transitions in detail. We show that the system crystallizes into an *rhcp* structure with different degrees of in-plane and out-of-plane stacking disorder that increase upon particle swelling. We further find that the melting process is distinctly different, where the system separates into two different crystal phases with different melting temperatures and interparticle interactions.

## 1. Introduction

Microgels feature an internal gel-like structure that can be highly swollen by solvent rendering them soft and deformable.[1,2] With their typical dimensions in the submicrometer range, microgels resemble many physical properties of classical colloids, while also behaviour common for macromolecules and surfactants are observed.[2-4] In addition, the microgel network can be finely tuned, for instance their softness can be modified through the degree of chemical cross-linking,[5-7] while a response to external stimuli, such as temperature, pH, or concentration gradients, can be reached by altering the chemical composition.[1,2,8] These complex interactions and their tunability render microgels as the ideal model system to study soft colloidal interactions and phase behaviour. In this role microgels have been used to study important fundamental phenomena, such as phase transitions, defect formation, as well as the glass transition or jamming in dense soft particle systems[9-21]. In particular, studies from the last few years addressed the microgel structure in densely packed systems, revealing that microgels at sufficiently high particle densities can experience different phenomena such as interpenetration[22-24] and/or deswelling.[22,25-29]

When microgels are prepared from poly-N-isopropylacrylamide (PNIPAM), the size and volume fraction of microgels can be controlled in-situ by temperature variations[30]. This makes PNIPAM microgels of particular interest for in situ investigations of phase transitions, such as crystallization and melting[10,15,17]. The temperature behaviour is related to the lower critical solution temperature (LCST) of PNIPAM in water below which polymer-solvent interactions are favoured. Above the LCST polymer-polymer interactions dominate leading to chain collapse into globules and results in the pronounced volume phase transition (VPT) behaviour. Below the VPT temperature (VPTT) PNIPAM microgels are highly swollen by water and thus possess large volumes. Surpassing the VPTT a strong deswelling is observed due the expulsion of water and the microgel volume can decrease by almost 90%.[30-32] During shrinkage a transition from soft repulsive to short-range attractive particle interactions is also observed for particles with weak to no electrostatic stabilization.[34,35] In contrast, a change from soft repulsive to less soft, electrostatic interactions is observed for microgels that possess more ionic groups.[30,32]

Importantly, the temperature responsive phase behaviour of PNIPAM microgels is still not completely understood. In particular, this is the case for high particle concentrations close to and above the VPTT where the exact particle interactions and their internal degrees of freedom become relevant. In a recent study by Bergman et al.[343] it was highlighted that upon approaching the VPTT the microgel interaction potential can be

[a.] Deutsches-Elektronen Synchrotron DESY, Notkestraße 85, 22607 Hamburg, Germany.
[b.] Moscow Institute of Physics and Technology (State University), Institutskiy Per. 9, 141701 Dolgoprudny, Moscow Region, Russia.
[c.] Department of Applied Chemistry, Graduate School of Advanced Science and Engineering, Hiroshima University, 1-4-1 Kagamiyama, Higashihiroshima 739-8527, Japan.
[d.] National Research Tomsk Polytechnic University (TPU), Lenin Avenue 30, 634050 Tomsk, Russia.
[e.] Heinrich-Heine-Universität Düsseldorf, Universitätsstraße 1, D-40225 Düsseldorf, Germany.
[f.] National Research Nuclear University MEPhI (Moscow Engineering Physics Institute), Kashirskoe shosse 31, 115409 Moscow, Russia.
[g.] Department of Applied Physics and Institute for Complex Molecular Systems, Eindhoven University of Technology, Groene Loper 19, 5612 AP Eindhoven, The Netherlands.
† corresponding author: ivan.vartaniants@desy.de
‡ corresponding author: j.m.meijer@tue.nl

best described by a multi-Hertzian model, taking into account repulsion from the higher cross-linked cores. One of the main reasons why the temperature response of PNIPAM microgels is hard to address, is the fact that upon close contact the microgels start to overlap and cannot be resolved individually. This explains why most optical (fluorescent) microscopy studies have focused on dilute systems[34] or crystalline systems in which the periodic order helps to resolve particle centers[10,15,17] and only intensive experimental optimization such as specific fluorescent labelling and super-resolution methods provide enough resolution to resolve the microgels in dense states[22]. Also, for scattering methods using e.g. neutrons or X-rays the microgels possess very little contrast and thus long measurement times are required. In addition, it has been shown that the microgel form factor significantly differs from the dilute, non-interacting state,[27] rendering the analysis of the structure factor of the dense state difficult.

The use of core-shell (CS) particles can circumvent several of these problems, as the cores can be labelled such that these can be easily detected, for instance with fluorescent dyes or high contrast materials, and thereby can provide information on the particle centre distributions.[35-38] Combined with the development of in situ techniques, their availability opens up the possibility to perform time-resolved in-situ studies during temperature-induced phase transitions, such as crystallization and melting, which is still not fully understood.[39] For in-situ studies small-angle X-ray scattering (SAXS) provides great resolution in space and time of and for this CS microgels with high electron density cores are desired. Ideally suited for this purpose are CS microgels with small, monodisperse gold nanoparticle cores that are accessible via seeded precipitation polymerization.[40,41] These particles are also of interest for several optical applications because gold nanoparticles feature localized surface plasmon resonances (LSPR) while the microgel shells can be used to control inter-particle spacing and assembly structures.[42,43] Periodic 2D lattices of these CS microgels were found to sustain surface lattice resonances (SLRs) as the result of plasmonic/diffractive coupling that arises when the inter-particle spacing is close to the LSPR.[44,45] The self-assembly into 3D crystals has been studied by UV-VIS spectroscopy and small-angle neutron scattering (SANS),[46] but structural changes induced by temperature were not explored yet.

Here, we investigate the phase behaviour of dilute and dense suspensions of Au-PNIPAM CS microgels with Ultra-Small Angle X-ray Scattering (USAXS). The gold core provides high X-ray scattering contrast while the particle interactions are governed by the microgel shell which makes this combination uniquely suited for in-situ investigations. We explore the phase transitions between crystalline and fluid-like states in response to both cooling and heating with a rate of 0.1 °C/min. We investigate the exact details of the processes using our recently developed Bragg peak analysis[47] and identify the crystal structure and structural changes during crystallization and melting. This allows us to identify the freezing and melting point but also reveals unexpected interparticle behaviour. In addition, we find that upon melting the system behaves differently compared to crystallization, showing the separation into three different crystallites consisting of two phases with distinctly different melting behaviour. Our results show that the combination of the CS microgels with USAXS opens up the possibility for detailed investigations of soft PNIPAM microgel phase behaviour upon changes in temperature and provides new fundamental insight into the nature of the phase transitions, also important for their application as functional materials.

## 2. Experimental section

### 2.1. Sample preparation

Temperature-sensitive CS microgels consisting of gold nanoparticle cores and chemically cross-linked microgel shells (PNIPAM) were synthesized following established protocols.[40,45] The obtained CS particles were found to contain an Au-core of the radius $R_{core}$ = 29.1 ± 4.2 nm and to have a hydrodynamic radius $R_h$, in the swollen state of $R_h$(20 °C) = 228.9 nm and in the collapsed state of $R_h$(50 °C) = 151.1 nm. The VPTT was determined to be at approximately 32.2 °C (see for details Supplementary Information, section S1 and Fig. S1). Two different dispersions of Au-PNIPAM particles with different concentrations, 0.5 wt. % and 12 wt. %, were prepared in deionized water (>18.2 MΩ·cm at 25 °C) and kept at these conditions by adding ~5 mg of ion exchange resin. The 12 wt. % dispersion showed upon visual inspection optical Bragg reflections at 20 °C and their absence at 50 °C indicating a phase transition. The effective volume fraction $\phi_{eff}$ of the samples at different temperatures was estimated from the CS particle volume via $R_h$ and the free volume based on interparticle spacing in the fully crystalline state of the 12 wt. % sample at 38 °C assuming an face-centered cubic *(fcc)* packing. We find at 20 °C for the 12 wt. % dispersion $\phi_{eff}$ = 0.60 and for the 0.5 wt. % dispersion $\phi_{eff}$ = 0.025 (see for details Supplementary Information, section S2). The dispersions (~20 µL) were placed into flat capillaries (4 × 0.2 × 50 mm$^3$, internal dimensions, Vitrocom) by employing a reduced pressure method. For the 12 wt. % sample the dispersion was heated to 50.0 °C prior to filling the capillary to reduce the dispersion viscosity. To prevent water evaporation during the experiment the open ends of the capillaries were flame sealed.

### 2.2. USAXS experiment

Ultra-small angle X-ray scattering was performed at the Coherence Applications Beamline P10 of the PETRA III synchrotron radiation facility at DESY, Hamburg. An X-ray beam with the photon energy *E* = 8.539 keV (wavelength *λ* = 0.145 nm) was cut down to the size of ~50 × 50 µm$^2$ on the sample by a slits system. A 2D detector EIGER X 4M (Dectris AG) with 2070 × 2167 pixels and a pixel size of 75 × 75 µm$^2$ was positioned 21.3 m behind the sample in USAXS geometry (Fig. 1). To avoid air absorption, an evacuated flight tube was placed between the sample and detector. The exposure time was selected to be 0.1 s to minimize radiation damage during the experiment. The sample capillaries were mounted in a copper sample holder which provided a uniform temperature distribution along the capillary. The holder had two small windows with a diameter of 1 × 4 mm$^2$ to allow X-rays to pass through the sample (See Supplementary information Fig. S2). Heating and cooling of the sample was performed by a Peltier element and circulating water bath. The temperature was measured by a thermocouple, which was in contact with the copper frame. A temperature controller adjusted the Peltier element to maintain a certain temperature with 0.001 °C stability. Measurements were performed in the temperature range between 20.0 °C and 50.0 °C.

## 3. Results

### 3.1. Core-shell particle and phase characterization

We first investigated the general properties of the CS microgels in the dilute state with 0.5 wt. % and $\phi_{eff}$(20 °C) = 0.025 between T = 25.0 °C and T = 50.0 °C. Examples of the 2D USAXS patterns measured in the fully collapsed state (T = 40.0 °C) and just slightly above the VPTT (T = 35.0 °C) are shown in Fig. 2a,b. The scattered intensity $I(q)$ is a product of the form factor $P_{cs}(q)$ of the CS particles and the structure factor $S(q)$ of the superlattice, $I(q) \propto P_{cs}(q)S(q)$. At this low volume fraction interference between scattering from different particles is negligible (i.e. $S(q) \approx 1$) and the resulting scattering represents solely the $P_{cs}(q)$ of the CS microgels. The radially averaged intensity profiles are shown in Fig. 2c (see also Supplementary Information, Fig. S3a,b). First of all, we note the large difference in scattering contrast between the gold core and the polymer shell that leads to two distinct features in the $P_{cs}(q)$, with a first minimum around $q \sim 30$ μm$^{-1}$ and a second minimum at $q \sim 160$ μm$^{-1}$, respectively. To extract the CS characteristics, the profiles were fitted with a core-shell model in which we accounted for the particle polydispersity by using a Gaussian size distribution (see Supplementary Information, section S4 for details of the fitting). The fitting was performed for each temperature and the evolution of the extracted parameters is shown in Supplementary Information Fig. S4. The core scattering contrast was fixed at $\Delta\rho_{core}$ = 4326 nm$^{-3}$ and the core radius was found to be $R_{core}$ = 25.8 ± 4.6 nm for all temperatures This Au-core size agrees well with $R_{core}$ = 29.1 ± 4.2 nm measured by TEM. In addition, the fits also confirm the size change of the PNIPAM shell with increasing temperature. We find that the total shell radius $R_{shell}$ decreases from $R_{shell}$(25 °C) = 192 ± 31 nm to $R_{shell}$(50 °C) = 162 ± 22 nm, while the shell scattering contrast $\Delta\rho_{shell}$ increases from $\Delta\rho_{shell}$(25 °C) = 16 nm$^{-3}$ up to $\Delta\rho_{shell}$(50 °C) = 25 nm$^{-3}$, confirming the collapse of the PNIPAM shell. This change in size agrees well with the observed change in the hydrodynamic radius $R_h$ from $R_h$(25 °C) = 220.8 nm to $R_h$(50 °C) = 151.1 nm. The discrepancy between $R_{shell}$ and $R_h$ is typically observed for microgels and can be explained by a fuzzy-sphere structure with lower cross-linking density and dangling ends in the outer region of the shell.[9] Here, this detail is ignored in the $P_{cs}(q)$ fit where a homogeneous density is assumed leading to a smaller $R_{shell}$.

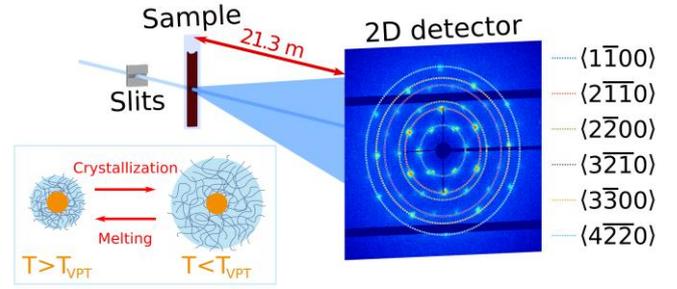

**Figure 1.** Scheme of the USAXS setup at the Coherence Applications Beamline P10 at PETRA III synchrotron storage ring. The beam was shaped with 50 × 50 μm2 slits before passing through the sample. The core-shell microgels were sealed in the glass capillaries. The scattering pattern was detected by an EIGER X 4M detector positioned 21.3 m behind the sample. Diffraction pattern shown in this figure was collected at T=35 °C. Families of Bragg peaks are indicated in the caption. For the crystalline sample this results in distinct Bragg peaks in the 2D USAXS pattern that are assigned to a random hexagonal close-packed (rhcp) crystal structure oriented along the [0001] axis. The inset on the bottom left shows a schematic representation of the swelling/deswelling behaviour of the PNIPAM shell of the CS microgels upon cooling/heating, resulting in a phase transition from a fluid to a crystal phase and vice versa.

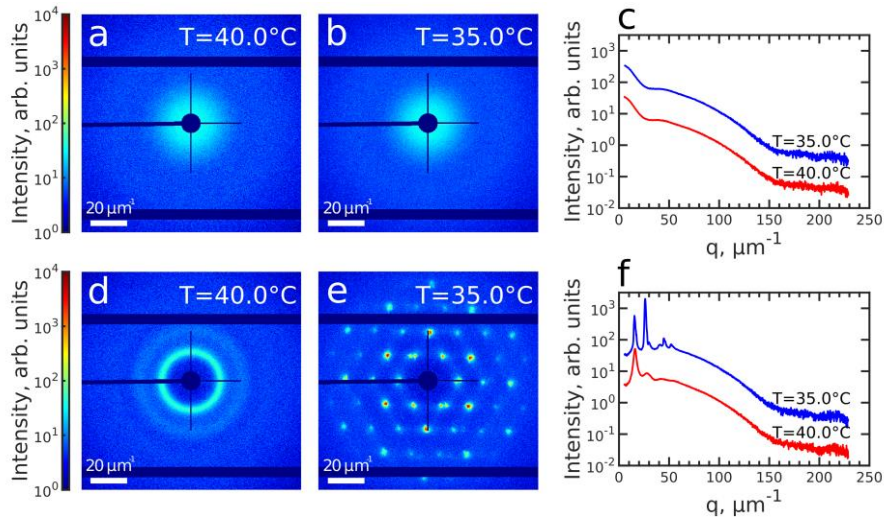

**Figure 2. (a,b)** Typical 2D-USAXS patterns of the dilute colloidal sample with 0.5 wt% at T = 40.0 °C (a) and T = 35.0 °C (b). **(c)** The corresponding radially averaged profiles of the scattered intensity. **(d,e)** Typical 2D-USAXS patterns of the densely packed colloidal sample with 12 wt% at T = 40.0 °C (d) and T = 35.0 °C (e). **(f)** The corresponding radially averaged profiles of the scattered intensity. In (c,f) the profiles of scattered intensity are offset by an order of magnitude for clarity.

Next, we investigated the high concentration sample with $\phi_{eff}$(20 °C) = 0.60 that showed a crystal to fluid phase transition between T = 25.0 °C to T = 50.0 °C, as evident from the appearance of optical Bragg reflections upon cooling. Examples of the 2D USAXS patterns in the collapsed state at T = 40.0 °C and close to the VPTT at T = 35.0 °C are shown in Fig. 2d,e, together with the radial averaged profiles shown in Fig. 2f. We assume the USAXS signal is dominated by scattering from the Au-cores due to two reasons: the higher scattering contrast of the Au-cores and the decrease in the contrast between the shells due to the dense packing of the CS microgels at high $\phi_{eff}$. Therefore, we attribute the main contribution to $I(q)$, and hence $S(q)$, to be originating from the Au-cores. At T = 40.0 °C, the 2D-USAXS pattern shows broad isotropic rings characteristic for scattering from a disordered fluid phase. At T = 35.0 °C, the 2D-USAXS pattern shows six prominent orders of narrow Bragg peaks originating from the CS microgels that have organized into a crystal lattice. The six-fold symmetry of the Bragg peaks can be attributed to a random hexagonal close-packed (*rhcp*) crystal lattice as indicated in Fig. 1 and will be discussed in details below.

### 3.2. *In-situ* characterization of crystallization

To investigate the crystallization process of the CS microgel system, the phase transition from fluid to crystalline state was followed *in-situ* with USAXS by applying continuous cooling around the temperature where the phase transition was observed. For this, the sample was first heated to T = 50 °C to allow the system to equilibrate in a fluid state, followed by a cooling step to T = 40 °C where the CS microgels already swell resulting in $\phi_{eff}$(40 °C) = 0.21 based on the $R_h$ but remain in a fluid-like state. Next, the system was continuously cooled with a rate of 0.1 °C/min and diffraction patterns were collected every 30 s from T = 39.0 °C to T = 35.0 °C giving a temperature resolution of 0.05 °C. At each temperature, radially averaged $I(q)$ profiles of the diffraction patterns were extracted. From these profiles, $S(q)$ can be obtained by dividing the measured intensity $I(q)$ by the fitted form factor $P_{cs}(q)$ from the dilute sample at each temperature step as $S(q) \propto I(q)/P_{cs}(q)$. The evolution of $S(q)$ as a function of temperature during the full cooling process is shown as an intensity map in Fig. 3a. At high temperatures (T > 38 °C, top part of Fig. 3a), $S(q)$ contains only broad features that can be attributed to the scattering from the isotropic fluid. At T = 38.2 °C the first sharp Bragg peaks start to appear, indicating the onset of crystallization. The size change of the PNIPAM shell at this particle concentration (12 wt. %) leads to a significant change in particle volume fraction from $\phi_{eff}$(40 °C) = 0.21 to $\phi_{eff}$(35 °C) = 0.30 in between these two temperatures $\phi_{eff}$ thus exceeds the freezing volume fraction $\phi_f$, i.e. $\phi_{eff} > \phi_f$, and results in the crystallization of the CS particles.

To follow the transitions of the fluid and crystalline phases separately, we extracted the isotropic structure factor $S_{iso}(q)$ by taking the average intensity on a ring at a $q$-value between the Bragg peaks and the crystal structure factor via $S_{xtal}(q) = S(q) - S_{iso}(q)$, which contains highly anisotropic features caused by the Bragg peaks of the crystalline phase. Fig. 3b and Fig. 3c show the distinctly different evolution of $S_{iso}(q)$ and $S_{xtal}(q)$ for the full temperature range, respectively. At high temperatures (T > 38.2 °C) $S_{iso}(q)$ shows only the fluid features that almost completely disappear upon crystallization of the sample (T < 37.6 °C). At T < 37 °C the small remaining intensity in $S_{iso}(q)$ comes only from the tails of the Bragg peaks.

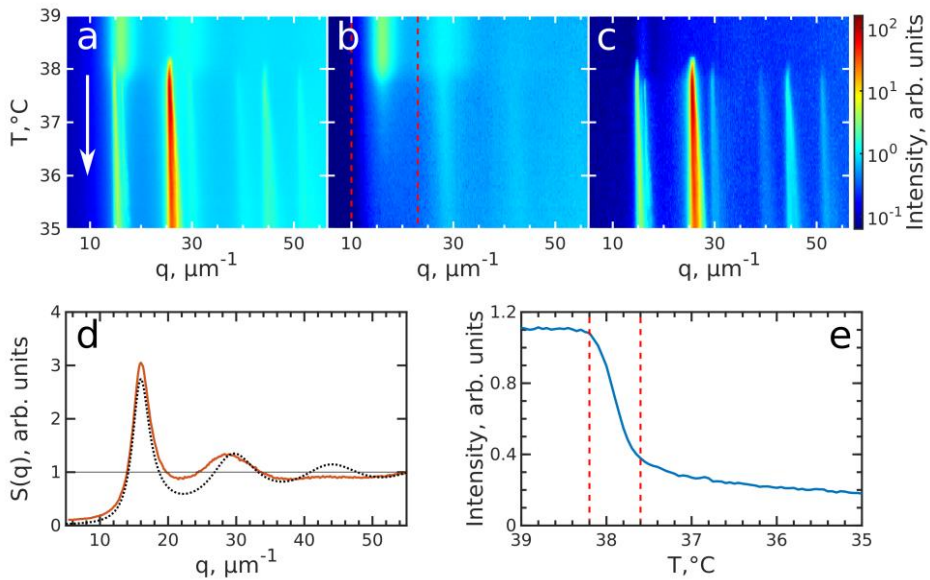

**Figure 3.** Evolution of crystallization of CS system at $\phi_{eff}$(20 °C) = 0.60 during cooling. The radially averaged intensity plots for different temperatures are stacked together in 2D maps as a function of scattering vector $q$ and temperature T for **(a)** full structure factor $S(q)$, **(b)** fluid structure factor $S_{iso}(q)$ (the intensity between the Bragg peaks), **(c)** crystal structure factor $S_{xtal}(q)$ (containing only the Bragg peaks). The white arrow in (a) indicates the direction of the experiment. **(d)** Fluid structure factor $S_{iso}(q)$ at T = 39.0 °C (red line) and the best Percus-Yevick hard sphere structure factor fit (black dotted line). **(e)** Integrated $S_{iso}(q)$ near the first fluid ring (in the range of $q$ = 10 – 23 µm$^{-1}$ as indicated by the red dashed lines in panel (b)). The vertical red dashed lines indicate the temperature range of active crystallization.

In contrast, at high temperatures (T > 38.2 °C) $S_{xtal}(q)$ shows very small traces of the first maximum of the fluid structure factor, while at T = 38.2 °C the appearance of the first Bragg peaks occurs that are followed by the appearance of higher order peaks between T = 38.2 – 38.0 °C which upon further cooling, continue to increase in intensity. We extracted the exact state of the fluid and the phase transition temperature from $S_{iso}(q)$. Fig. 3d shows $S_{iso}(q)$ at $T$ = 39.0 °C where a broad first maximum from the fluid phase can be seen. We fitted the $S_{iso}(q)$ with the Percus-Yevick hard sphere model $S_{PY}(q)$ (see Supplementary Information, section S5 for details of the fitting).[48] The best fit for $S_{iso}(q)$ with a hard sphere radius of $R_{PY}$ = 216 ± 1 nm and a volume fraction of $\phi_{PY}$ = 0.47 ± 0.03 is also shown in Fig. 3d. The obtained $R_{PY}$ is larger than the CS size of $R_h$(39 °C) = 162.7 nm and can be explained by the charged characteristics of our microgels with a zeta-potential of ζ ≈ -30 mV that dominates the particle interactions in the collapsed state (see for details Supplementary Information Fig. S1). The surface charges result in long-range electrostatic repulsion between the CS particles under the deionized conditions that leads to a Debye length of $\kappa^{-1}$ ~ 100 nm. The high volume fraction $\phi_{PY}$ = 0.47 indicates the system is indeed showing signatures of a fluid close to the hard sphere freezing volume fraction $\phi_{f-HS}$ = 0.494. From the fluid structure factor intensity $S_{iso}(q)$ the onset and end of the full crystallization process were determined. Fig. 3e shows the integrated value of $S_{iso}(q)$ around the first maximum in the range of 10 – 23 μm$^{-1}$ (red dashed lines in Fig. 3b). The intensity remains constant up to T = 38.2 °C and then starts to drop significantly, which coincides with the appearance of the Bragg peaks in $S_{xtal}(q)$. Clearly, this temperature is the starting point for crystallization of the sample and at this point $\phi_{eff} = \phi_f$ = 0.23, as a result of the small increase of the CS particle size ($R_h$(38 °C) = 165.8 nm). The major drop in intensity occurs between T = 38.2 – 37.6 °C and indicates the crystallization of the major part of the system during this small temperature and time window (0.6 °C, 6 min), which we will refer to as the 'active crystallization' regime. We note that further cooling still leads to a small decrease in the $S_{iso}(q)$ intensity which can be caused by crystallization of residual amounts of the fluid phase as well as potential annealing of crystalline defects that would decrease the intensity of the Bragg peak tails (further discussed below).

To investigate the crystallization process in more detail, we performed Bragg peak analysis on the peaks visible in the 2D USAXS patterns.[47] There are six prominent orders of Bragg peaks present in $S_{xtal}(q)$ at $q/q_1 = 1, \sqrt{3}, 2, \sqrt{7}, 3, 2\sqrt{3}$ with respect to the first order peak at $q_1 \approx 15$ μm$^{-1}$. These Bragg peaks indicate the presence of a single crystal domain with a random hexagonal close-packed (*rhcp*) structure aligned with its hexagonal close-packed planes parallel to the capillary walls and, hence, perpendicular to the X-ray beam. The *rhcp* structure is typically found for colloidal spheres, as the spheres pack into close packed hexagonal planes while the stacking sequence of the planes is random, leading to alternating *fcc* and hexagonal close-packed (*hcp*) crystal structures.[14, 49-52] The Bragg peaks can be identified as the hexagonal close packed ⟨1$\bar{1}$00⟩, ⟨2$\bar{1}\bar{1}$0⟩, ⟨2$\bar{2}$00⟩, ⟨3$\bar{2}\bar{1}$0⟩, ⟨3$\bar{3}$00⟩ and ⟨4$\bar{2}\bar{2}$0⟩ families, respectively (see Fig. 1). We do note that there are two additional peaks of lower intensity next to the ⟨1$\bar{1}$00⟩ peaks in the pattern. We believe that these peaks originate from another crystal grain and were therefore excluded from the further analysis.

The Bragg peak analysis was done by fitting the peaks with a 2D Gaussian function (see Supplementary Information, section S6 for details). Each diffraction pattern was interpolated into a polar ($q,\varphi$)-coordinate frame and divided by the corresponding single particle form-factor. Each Bragg peak was fitted separately with a 2D Gaussian function in the polar coordinates. The following fitting parameters were extracted: the peak intensity, the $q$-position of the center of the peak, and the full widths at half maximum (FWHMs) in radial and azimuthal directions. Finally, the obtained values were averaged for each Bragg peak family with the error bars representing the standard deviation within each family.

The evolution of the integrated Bragg peak intensities for each Bragg peak family over the full investigated temperature range is shown in Fig. 4a. The first peaks to appear are the brightest ⟨2$\bar{1}\bar{1}$0⟩ family peaks at T = 38.25 °C, confirming again that at this temperature the crystallization starts. Upon further cooling, higher order Bragg peaks appear, with the last set of peaks belonging to the ⟨3$\bar{2}\bar{1}$0⟩ family, which also possesses the lowest intensity, at T = 38.05 °C. The intensity of all peaks rapidly increases from the moment they appear until the intensity increase significantly slows down for temperatures approaching T = 37.6 °C. Further cooling only leads to a minor rise off all intensities. The appearance and the rise in intensity of the peaks up to T = 37.6 °C, indicates the growth of a crystal nuclei from the fluid and the increasing long-range order of the crystal grain.

The evolution of the peaks position with respect to the initial $q$-values, $q_0$, is shown in Fig. 4b. During the initial rapid crystallization, the peak positions stay quite stable or even slightly decrease. However, after T = 37.6 °C, when most of the sample has crystallized, all Bragg peaks start moving towards higher q-values, indicating that the lattice spacing decreases. Surprisingly, the increase rate varies for different Bragg peak families. For the most intense ⟨2$\bar{1}\bar{1}$0⟩ peaks, the $q$-value increases only by ~2% from the start of crystallization to the final temperature $T$ = 35.0 °C, while that of the ⟨1$\bar{1}$00⟩ peak increases up to ~5%. All other Bragg peaks move with rates between these two extrema, although ⟨2$\bar{2}$00⟩ and ⟨3$\bar{2}\bar{1}$0⟩ are closer to the rate of ⟨1$\bar{1}$00⟩, and ⟨3$\bar{3}$00⟩ and ⟨4$\bar{2}\bar{2}$0⟩ are closer to ⟨2$\bar{1}\bar{1}$0⟩. Interestingly, these two groups of peaks have different origins. In reciprocal space the stacking disorder of the planes in the *rhcp* structure leads to the appearance of stacking-dependent Bragg rods along the direction normal to the close packed planes, which in this case is parallel to the X-ray beam. Here, the ⟨1$\bar{1}$00⟩, ⟨2$\bar{2}$00⟩ and ⟨3$\bar{2}\bar{1}$0⟩ can be identified as stacking-dependent peaks, while the ⟨2$\bar{1}\bar{1}$0⟩, ⟨3$\bar{3}$00⟩ and ⟨4$\bar{2}\bar{2}$0⟩ are stacking-independent peaks. The difference in their $q$-value evolution seems to indicate that there are differences in how the crystal grows and how the defect structure develops in the in- and out-of-plane direction of the crystal grain.

From the peak positions of all Bragg peaks, we can calculate the average *hcp* unit cell parameter $a$ as shown in Fig. 4c. During the active crystallization $a$ is almost constant, only increasing slightly from $a = 489 \pm 1$ nm to $a = 491 \pm 1$ nm. However, further cooling leads to a decrease to $a = 476 \pm 5$ nm. The interparticle spacing upon crystallization is larger than $2R_h(38\,°C) = 331.7$ nm and its decrease with further cooling is contradictory to the swelling of the PNIPAM shells to $2R_h(35\,°C) = 362.3$ nm (see Supplementary Information, Fig. S1c). Both discrepancies seem to be caused by the electrostatic interactions between the CS particles that as mentioned above lead to a long-range interparticle interaction and hence an earlier onset of crystallization. It has been shown for ionic microgels that at a fixed temperature an increase in number density (and thus $\phi_{eff}$) results in a decrease in interparticle spacing.[11,53] Moreover, at high enough particle concentration the overlap of the counterion clouds can even lead to deswelling of the microgels.[25] However, in our case the situation might be even more complex as we find apparent microgel charge changes in dependence on the swelling state as evidenced by the different zeta potentials, i.e. $\zeta(38\,°C) = -25.6$ mV and $\zeta(35\,°C) = -18.5$ mV (see for details Supplementary Information Fig. S1). We do note that these values were obtained in the dilute system and thus might not reflect the dense system case. Clearly, the decrease in interparticle spacing upon cooling is the result of a complex change in interparticle interactions of the microgels, and remains a topic that is still not fully understood.[2,33]

We can further extract information about the distortions caused by strain in the crystal lattice by performing Williamson-Hall analysis of the Bragg peaks.[54] For this we determined the FWHM of each Bragg peak $w_q$ and $w_\varphi$ in radial and azimuthal direction, respectively. Fig. 4d,g show $w_q$ and $w_\varphi$ averaged for each Bragg peak family for the full temperature range. While each family has different absolute values their overall trends of $w_q$ and $w_\varphi$ are quite similar. There is, however, a clear difference between the trends in $w_q$ and $w_\varphi$. While $w_q$ continuously grows throughout the whole cooling process indicating continuously growing strain in the crystal lattice, $w_\varphi$ only increases during the active crystallization between $T = 38.2\,°C - 37.6\,°C$ and then remains constant showing that after crystallization the strain in this directions does not evolve. Using the Williamson-Hall equation[55]

$$w_{q,\varphi}^2(q) = g_{q,\varphi}^2 q^2 + \left(\frac{2\pi}{L_{q,\varphi}}\right)^2, \quad (1)$$

where $w_{q,\varphi}(q)$ is the FWHM of a Bragg peak at position $q$, $g_{q,\varphi}$ the lattice distortion and $L_{q,\varphi}$ the size of coherently scattering domains; subscripts $q$ and $\varphi$ denote radial and azimuthal directions, respectively. Fig. 4e,h show $w^2_q$ and $w^2_\varphi$ for all Bragg peak families at $T = 36.0\,°C$ where the sample has fully crystalized. In contrast to the prediction of the Williamson-Hall equation, the points of the different Bragg peak families do not fall on a single straight line but separate again into stacking-dependent peaks ($\langle 1\bar{1}00\rangle$, $\langle 2\bar{2}00\rangle$ and $\langle 3\bar{2}\bar{1}0\rangle$) and stacking-independent ones ($\langle 2\bar{1}\bar{1}0\rangle$, $\langle 3\bar{3}00\rangle$ and $\langle 4\bar{2}\bar{2}0\rangle$). This separation agrees well with previously observed differences in the FWHMs of stacking-dependent and stacking-independent Bragg peaks for a similar colloidal system with *rhcp* packing.[50] The difference in the FWHMs is caused by the presence of additional in-plane stacking disorder, which means that the hexagonal planes consist of islands with different lateral positions with characteristic line defects in between them leading to a lower degree of order in these in-plane directions.[49] In addition, for soft colloidal crystals different types of defects consisting of combined in-plane and out-of-plane stacking disorder, i.e. partial dislocations, have been observed and would lead to a similar effect.[56] Therefore, we fitted the stacking-dependent and stacking-independent Bragg peaks separately and extracted the lattice distortions $g_q$ and $g_\varphi$ over the full temperature range, as shown in Fig. 4f and 4i.

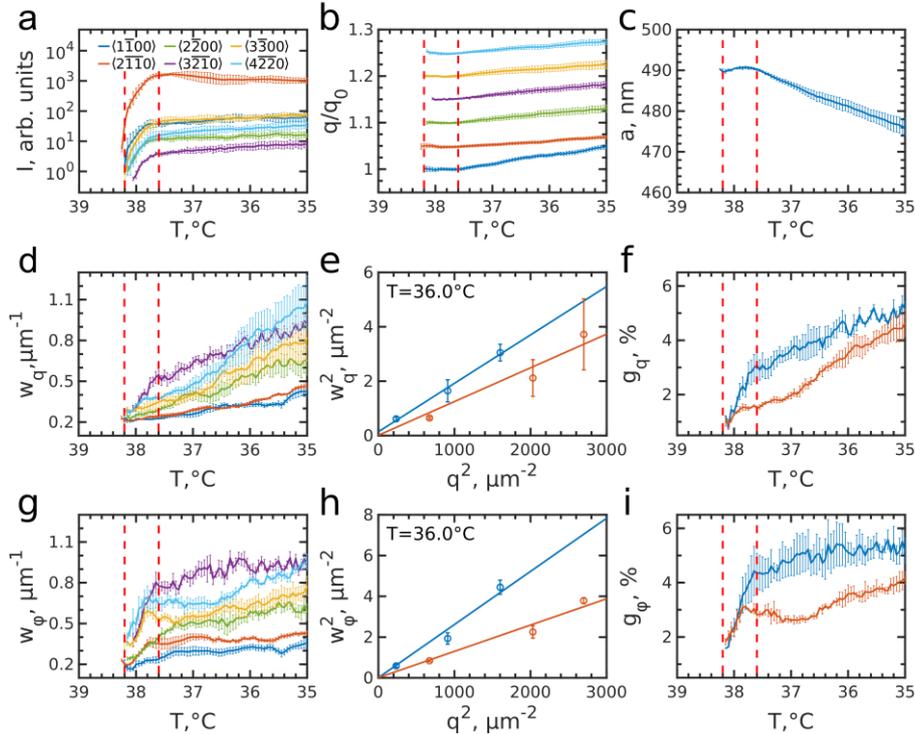

**Figure 4.** Evolution of the Bragg peaks and crystal parameters during cooling. In all panels the vertical red dashed lines indicate the temperature range of active crystallization. **(a)** Integrated intensity, **(b)** $q$-position of the peaks in respect to the first registered q-value of the peak, $q_0$ (plots are offset by 0.05 for clarity). The error bars are standard deviations between the peaks of the same family. **(c)** Evolution of the lattice parameter, $a$, averaged over all observed Bragg peaks. **(d,g)** The size of the Bragg peaks (FWHM) in radial (d) and azimuthal (g) directions. The error bars are standard deviations between the peaks of the same family. **(e,h)** Examples of the Williamson-Hall plots for FWHMs of the subpeaks in radial (e) and azimuthal (h) directions at $T = 36.0\,°C$. Points are experimental values for the stacking-dependent (blue) and stacking-independent (red) peaks and straight lines are the best fit for each group. **(f,i)** Evolution of the radial (f) and angular (i) lattice distortions extracted by the Williamson-Hall method from the stacking-dependent (blue) and stacking-independent (red) peaks.

We find that in the radial direction, the lattice distortions for both stacking types are initially the same with $g_q \approx 1\%$, indicating the initial crystal grain experiences little strain. As the active crystallization proceeds, the distortions start to increase with a higher rate for the stacking-dependent peaks than for the stacking-independent Bragg peaks. Upon further cooling to T = 35.0 °C, the radial distortions reached values of $g_q \approx 5.5\%$ and $g_q \approx 4.5\%$ for stacking-dependent and stacking-independent peaks, respectively. This larger distortion for the stacking-dependent peaks is expected since the in-plane stacking disorder leads to effectively smaller crystal domains. The continuous character of the increased distortion seems to be related to the continuous swelling of the CS microgels and accompanying softer interparticle interactions, similar to the decrease in interparticle spacing observed from the Bragg peak positions. The swelling leads to increasing strain in the crystals and thus will lead to larger distortions within the crystal planes.[56]

In the azimuthal direction, the crystallization is characterized by a fast rise of the angular distortion $g_\varphi$ in the active crystallization regime from 1.5% to 4% for stacking-dependent and from 1.5% to 3% for stacking-independent peaks, respectively (see Fig. 4i). This behaviour can be explained by the misorientation of the outsides of the growing crystal with respect to the nuclei orientation during the active crystallization stage. After the active crystallization, the distortions only increase slightly during further cooling reaching values of $g_\varphi \approx 5\%$ and 4%, respectively. Clearly, once the full scattering volume has crystallized, the additional strain from the particle swelling does not lead to strong reorientation of the crystal planes.

### 3.3. In-situ characterization of melting

After having analysed the crystallization process in detail, we now turn to the melting process induced by slowly heating the crystalline sample. We note that after the cooling measurement the sample was cooled further to T = 20 °C and equilibrated for 5 min. Next, the melting was followed from T = 35.0 °C to T = 43.0 °C with the same rate of 0.1 °C/min and at the same sample position as at the end of cooling. During the heating process, we observed that the intensity of the 6-fold Bragg peaks decreased and that the shape of the peaks changed.

Again, we identify the onset of melting by investigating the structure factors shown in Fig. 5. The evolution of $S(q)$, $S_{iso}(q)$ and $S_{xtal}(q)$ is shown in Fig. 5a-c. Fig. 5d shows yhe $S_{iso}(q)$ of the fluid state at T = 43.0 °C together with the best fit with the Percus-Yevick $S_{PY}(q)$ for a fluid with $\phi_{PY} = 0.42 \pm 0.01$ and $R_{PY} = 209 \pm 4$ nm. Fig. 5e shows integrated $S_{iso}(q)$ over the full temperature range.

The transition from a crystal to a liquid can be observed clearly by the appearance of the fluid peak in $S_{iso}(q)$ and the disappearance of the Bragg peaks in $S_{xtal}(q)$. In the temperature range of T = 37.0 – 38.2 °C a strong rise in the intensity of $S_{iso}(q)$ occurs that coincides with the most significant drop in the Bragg peak intensities, indicating the onset and subsequent melting of a main part of the crystalline phase. The lower T = 37.0 °C for the start of melting shows that the melting transition occurs at the higher volume fraction $\phi_{eff}$ = 0.24 than the freezing transition. With further heating, $S_{iso}(q)$ intensity only increases gradually, indicating that a large part of the sample still remains crystalline. This is also seen in the $S_{xtal}(q)$ peaks that decrease in intensity, but remain present up to T ≈ 41.0 °C. Finally, at T = 41.8 °C the Bragg peak intensity decreases and the peaks completely disappear at T = 42.3 °C while the liquid structure factor $S_{iso}(q)$ reaches its maximum intensity, indicating the full sample has returned to a fluid state. These PY values again agree with a collapsed PNIPAM shell state but are slightly larger than $R_h$(43 °C) = 155.5 nm, again attributed to electrostatic repulsions between the particles. Interestingly, in $S_{xtal}(q)$ the Bragg peaks appear much sharper between T = 38.2 °C and T = 41 °C, which is counterintuitive for a "normal" crystal melting. Typically melting is associated with an increase in lattice distortion and a decrease in the range of structural order that would lead to broadening of the Bragg peaks. After a more careful analysis, we found that the previously single crystal Bragg peak splits into distinctly different sets of Bragg peaks upon melting and that these peaks show different behaviour. We note that for T > 42 °C suddenly different Bragg peaks appear in $S_{xtal}(q)$ which coincides with a drop in $S_{iso}(q)$. We believe this moment indicates the moment the crystal grain fully breaks up and rotates or possibly the drift of another crystallite into the X-ray beam.

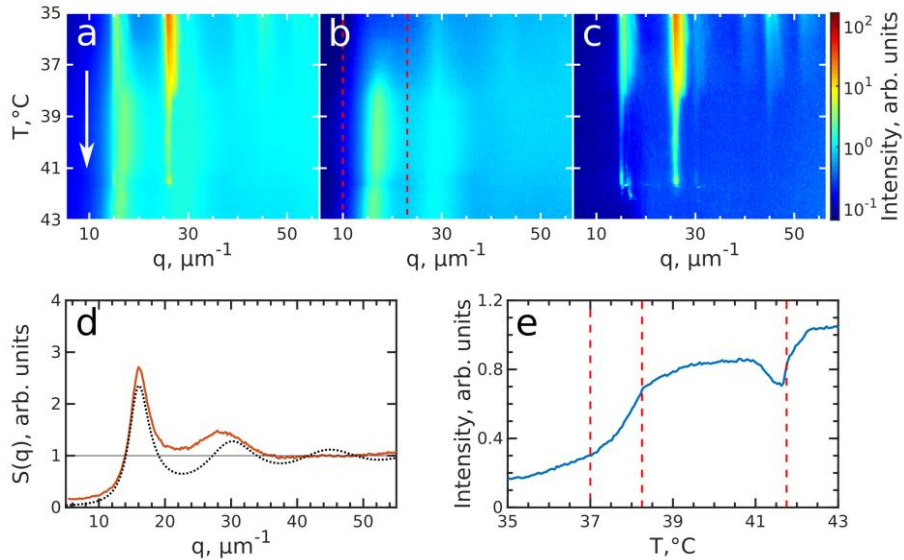

**Figure 5.** Evolution of melting of CS system with $\phi_{eff}(20 °C) = 0.60$ during heating from T = 35 °C to T = 43 °C. The radially averaged intensity plots for different temperatures are stacked together in 2D maps for (**a**) full structure factor $S(q)$, (**b**) fluid structure factor $S_{iso}(q)$ (the intensity between the Bragg peaks), (**c**) crystal structure factor $S_{xtal}(q)$ (containing only the Bragg peaks). The white arrow in (a) indicates the direction of the experiment. Note the reverse temperature scale compared to Fig. 3. (**d**) Measured $S_{iso}(q)$ at final temperature T = 43.0 °C (red line) and the best fit by the Percus-Yevick hard sphere structure factor (black dotted line). (**e**) Integrated $S_{iso}(q)$ near the first fluid ring (in the range of $q = 10 – 23$ μm$^{-1}$ indicated by the red dashed lines in panel (b)). The first two vertical red dashed lines indicate the temperature range of active melting and the last the final stage of melting.

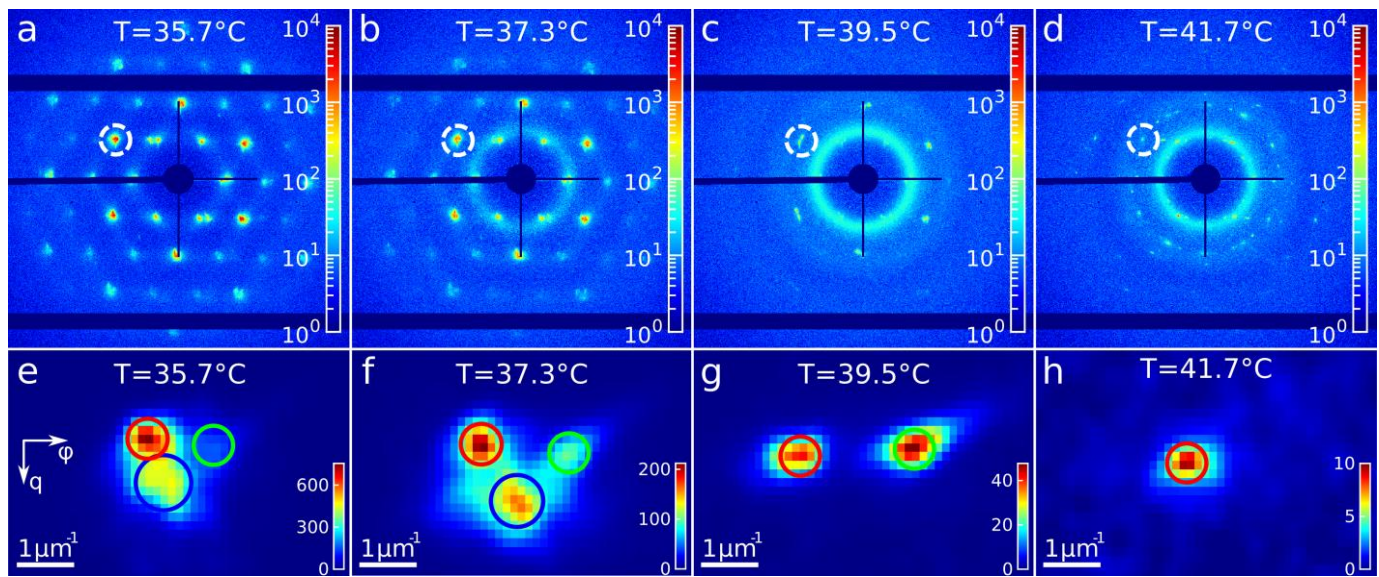

**Figure 6.** Evolution of the Bragg peaks during heating. **(a-d)** Examples of the 2D-USAXS patterns collected during heating at different temperatures. **(e-h)** Areas of the diffraction patterns showing thee subpeaks of the $2\bar{1}\bar{1}0$ Bragg peak, indicated in (a-d) by the dashed circle, at different temperatures.

To get insight into the unexpected behaviour of the sample during melting, we examined the 2D USAXS patterns in more detail. Fig. 6a-d shows the patterns at different temperatures and Fig. 6e-h shows magnified parts around one of the peaks from the brightest $\langle 2\bar{1}\bar{1}0 \rangle$ family. It can be clearly seen that the Bragg peak splits into three subpeaks already at the start of the measurement at T = 35.7 °C. Their appearance implies that during melting the previously single crystalline structure separates into three crystallites surrounded by a fluid phase, which can be explained by the onset of melting at defects and grain boundaries position.[10] The subpeaks corresponding to these crystallites are denoted "blue", "red" and "green" as indicated by the circles in Fig. 6e-h. We performed Bragg peak analysis of these three subpeaks and the evolution of the average extracted peak intensity, q-position and φ-position that are shown in Fig. 7. From the different parameters, it is clear that the "blue" crystallite behaves differently compared to the "green" and "red" crystallites. First, the "blue" peak has a higher intensity than the other two (Fig. 7a). Second, although heating up to T = 37 °C results in a decrease in the intensities of all three peaks, further heating results in a rapid decrease and disappearing at T = 38.2 °C of the "blue" subpeak, while the "red" and "green" subpeaks keep decreasing in intensity and only fully disappear at T = 41.8 °C. We note that we can exclude sedimentation of the crystallites at this stage of the heating process as the effective volume fraction $\phi_{eff} > \phi_f$. Third, the "green" and "red" subpeaks move apart from each other in azimuthal direction by about four degrees while the blue subpeaks do not move (Fig. 7b). Finally, during heating the q-position of the "blue" peak increases significantly while those of the red and green subpeaks stay relatively constant (Fig. 7c). From this analysis, it is clear that the "blue" crystallite comprises the bulk of the system since it shows the reverse behaviour with full melting at the same temperature as where bulk crystallization started. Therefore, the behaviour of this "blue" crystallite is driven by the CS particle size change and corresponding change in the effective volume fraction $\phi_{eff}$. For the "green" and "red" crystallites we conclude that these comprise a small part of the scattering volume and since these crystallites remain present after melting of the bulk of the system, it can explain the occurrence of drift and orientational changes of the crystallites.

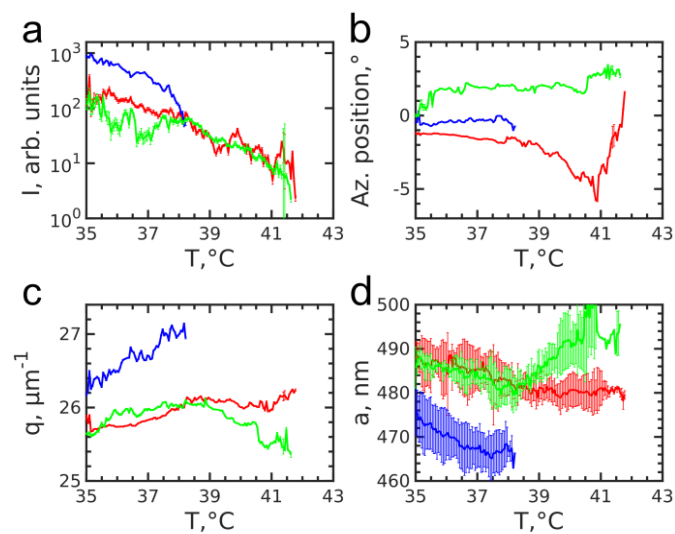

**Figure 7.** Evolution of the extracted parameters of the subpeaks of the $2\bar{1}\bar{1}0$ Bragg peak during melting: **(a)** integrated intensities, **(b)** azimuthal positions and **(c)** q-values. **(d)** Evolution of the lattice parameters for each superlattice crystallite. The lattice parameters are averaged over 5 orders of each subpeak.

Next, we calculated the average unit cell parameter for the crystallites from the average values of all orders of each subpeak (Fig. 7d). The subpeaks corresponding to the same crystallite in each Bragg peak family were identified thanks to their similar behaviour in radial and azimuthal directions. The extracted parameters of the separate Bragg peak analysis are shown in the Supplementary Information in Fig. S7-S10, where we note that the $\langle 3\overline{2}10 \rangle$ family was excluded due to its low intensity. For the "blue" crystallite we find the lattice parameter value $a = 475 \pm 5$ nm at T = 35 °C that decreases to $a = 468 \pm 5$ nm just before melting at T = 37 °C. This initial lattice spacing corresponds to the end value of the crystallization process and the decrease in lattice spacing agrees with the collapsing of the CS size. For the "green" and "red" crystallites we find $a = 489 \pm 3$ nm at T = 35 °C, which is 14 nm larger than at the end of crystallization. In addition, for the "red" crystallite $a$ continuously decreases to $a = 480 \pm 3$ nm at T = 38.2 °C and then stays constant up to melting, while for the "green" a decrease occurs to $a = 482 \pm 3$ nm at T = 38.2 °C after which it increases again up to $a = 492 \pm 4$ nm, exceeding thus the initial value. This behaviour of the "green" and "red" crystallite is surprising as it is not in-line with the expected collapse of the PNIPAM shell.

Based on the separation into three crystallites and their distinct differences in behaviour, we conclude that in the CS system two different states are present during melting. Based on the USAXS patterns alone it is difficult to determine the exact location of the crystals but we speculate that the distinction comes from the bulk crystal and two wall crystallites. Our reasoning is that since the "blue" crystallite shows the expected melting behaviour compared to the system upon crystallization, this crystallite comprises the bulk of the system and the melting transition is driven by the change in CS size and the corresponding change in $\phi_{eff}$. The similarities in behaviour of the "green" and "red" crystallites indicate that these crystals might be two crystalline domains formed on the capillary walls, as observed in other charged particle systems[57]. A temperature gradient close to the walls induced by the short cooling to T = 20 °C of the sample before the heating measurement will lead to increased swelling of the CS microgel shell ($R_h(20 °C) = 228.9$ nm) and could explain the larger lattice spacing in these crystals close to the wall. In addition, the swelling would lead to entanglement of the outer chains of the PNIPAM shell, and thus lead to an early separation into two states, while also explaining the delayed melting of the crystallites even below $\phi_f$. However, real space investigations, such as polarization microscopy[38,57], will be needed to confirm this melting process.

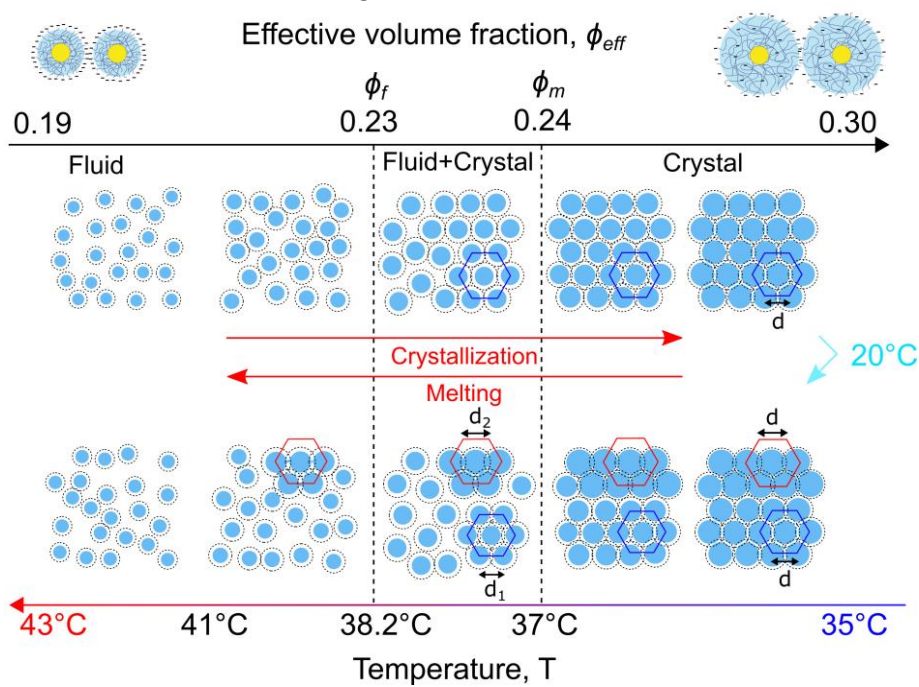

**Figure 8.** Schematic representation of the observed *in-situ* phase transitions from a fluid to a crystal and vice versa in the gold-PNIPAM CS system. Due to the swelling and collapsing of the PNIPAM shell in response to cooling and heating the CS microgel size and subsequently the effective volume fraction $\phi_{eff}$ in the system changes driving the phase transitions.

## 4. Discussion

A schematic representation of the full phase behaviour observed in the dense CS particle system upon cooling and heating with 0.1 °C/min is presented in Fig. 8. At high temperatures T > 39 °C the system is in a fluid state with the effective volume fraction $\phi_{eff} < \phi_f$. Upon cooling the charged CS particles start to swell and at T = 38.2 °C the effective volume fraction $\phi_{eff}$ increases above the freezing point to $\phi_{eff} > \phi_f$, causing the formation of a crystal nucleus that subsequently grows from the fluid. Upon a further decrease in temperature T < 37.8 °C, the particle swelling increases further, thus increasing $\phi_{eff}$ and due to their charged and soft nature the inter-particle spacing decreases. Upon cooling the system to T = 20 °C for 5 min the CS particles close to the capillary wall swell even more and the dangling ends in the outer microgel periphery can interpenetrate. Subsequently, by heating the system again, the microgel collapse lowering the effective volume fraction. At this stage the system starts to separate into two distinct crystal states, the bulk crystal and the wall crystals. Below the melting point, $\phi_{eff} < \phi_m$, the bulk of the system starts to melt and form a fluid phase, while the wall crystals respond much slower due to possible entanglement of the outer polymer chains of the microgels. Finally, only by heating to T > 41.5 °C the full system disperses again and a fluid phase is obtained.

## 5. Conclusions

We have investigated the *in-situ* crystallization and melting of CS microgels that contain high contrast Au cores and thermoresponsive PNIPAM shells using USAXS. The phase transitions were induced either by cooling or by heating with rates of 0.1 °C/min in a densely packed suspension. We have found that the Au-core contribution dominates the scattered intensity due to their high electron density and provides sharp contrast between the CS form factor $P_{cs}(q)$ and structure factor $S(q)$, while the temperature response of the PNIPAM shell can still be monitored. We further found that the systems behaviour upon crystallization and melting is quite different and rather complex. Upon cooling, due to the increase in CS size, the dense system readily crystallizes into a single crystalline structure. By performing Bragg peak analysis, we revealed that the system forms an *rhcp* crystal structure and that in-plane and out-of-plane stacking disorder occurs and develops differently. Upon heating the crystalline sample melts again but we find that different, smaller crystallites are formed. We attribute these differences to the presence of two different crystalline phases, a bulk phase and attached to the capillary wall crystals.

The combination of CS microgels containing Au cores and PNIPAM shells with USAXS and Bragg peak analysis employed in the current study provides a means to investigate the bulk behaviour of microgel systems upon temperature changes which have been limited so far due to the low scattering contrast. Our investigations therefore open up ways to address how the bulk system response ties in with the local microgel response for different particle concentrations as well as the influence of different cooling and heating rates. Such investigations can address fundamental questions regarding crystallization, melting, jamming and the glass transition, while at the same time these can provide crucial insights for potential applications of such microgels as (multi-)functional materials in various fields.

## Author Contributions

J.M.M., M.K and I.A.V. conceptualized the research. J.M.M. and M.K. performed sample preparation. D.L., D.A., J.S., F.W., S.L., M.S., M.K., I.A.V. and J.M.M. performed synchrotron experiments and data acquisition. D.L., N.M and S.D. performed data analysis. D.L., N.M., S.L., M.K., I.A.V. and J.M.M. interpreted results and wrote the manuscript. All authors read and agreed on the final text of the paper.

## Conflicts of interest

There are no conflicts of interest to declare.

## Acknowledgements

We acknowledge DESY (Hamburg, Germany), a member of the Helmholtz Association HGF, for the provision of experimental facilities. Parts of this research were carried out at PETRA III synchrotron facility and we would like to thank all the beamline staff for assistance in using Coherence Application beamline P10. We would like to thank Andrei Petukhov for valuable discussions. We further acknowledge Astrid Rauh for the particle synthesis, Max Schelling for the DLS measurements and Sanam Foroutanparsa for the TEM images. J.M.M. acknowledges financial support from the Netherlands Organization for Scientific Research (NWO) (016.Veni.192.119). M.K. acknowledges the German Research Foundation (DFG) for funding under grant KA3880/6-1. D.L., N.M., D.A., S.L. and I.A.V. acknowledge the Helmholtz Associations Initiative Networking Fund (Grant No. HRSF-0002) and the Russian Science Foundation (Grant No. 18–41-06001). I.A.V. acknowledges the financial support of the Russian Federation represented by the Ministry of Science and Higher Education of the Russian Federation (Agreement No. 075-15-2021-1352). S.L. acknowledges Competitiveness Enhancement Program Grant of Tomsk Polytechnic University and the Governmental program "Science", project No. FSWW-2020-0014.

# In-situ Characterization of Crystallization and Melting of Soft Thermoresponsive Microgels by Small-Angle X-ray Scattering

## Supplementary Information

Dmitry Lapkin[1], Nastasia Mukharamova[1], Dameli Assalauova[1], Svetlana Dubinina[1,2], Jens Stellhorn[1,3], Fabian Westermeier[1], Sergey Lazarev[1,4], Michael Sprung[1], Matthias Karg[5,*], Ivan A. Vartanyants[1,6,*], and Janne-Mieke Meijer[7,*]

[1]Deutsches-Elektronen Synchrotron DESY, Notkestrasse 85, 22607 Hamburg, Germany
[2]Moscow Institute of Physics and Technology (State University), Institutskiy Per. 9, 141701 Dolgoprudny, Moscow Region, Russia
[3]Department of Applied Chemistry, Graduate School of Advanced Science and Engineering, Hiroshima University, 1-4-1 Kagamiyama, Higashihiroshima 739-8527, Japan
[4]National Research Tomsk Polytechnic University (TPU), Lenin Avenue 30, 634050 Tomsk, Russia
[5]Heinrich-Heine-Universität Düsseldorf, Universitätsstraße 1, D-40225 Düsseldorf, Germany
[6]National Research Nuclear University MEPhI (Moscow Engineering Physics Institute), Kashirskoe shosse 31, 115409 Moscow, Russia
[7]Department of Applied Physics and Institute for Complex Molecular Systems, Eindhoven University of Technology, Groene Loper 19, 5612 AP Eindhoven, The Netherlands

*Corresponding authors: ivan.vartaniants@desy.de, j.m.meijer@tue.nl

## S1. Core-shell particle characterization

The size of the gold core of the core-shell (CS) microgel particles was determined with Transmission Electron Microscopy (TEM, FEI Tecnai 20 (type Sphera)). Fig. S1a shows a representative TEM images of the core-shell particles, where the dense gold core (black centres) can be clearly distinguished from the hydrogel shell (light grey corona). Note that due to drying effects, the shells have collapsed into a pancake shape resulting in different sizes and the gold core to appear off-centre. Some of the microgel particles were found to not contain a gold core but these constitute less than 0.1% of all particles. Using ImageJ particle analysis the gold core size was determined from the area of the cores in the TEM images for more than 1000 particles, assuming a spherical shape (Fig. S1b). The average gold core radius was found to be $R_{core}$ = 29.1 ± 4.2 nm.

The size of the CS microgels dispersed in water over a temperature range of T = 15 – 50 °C was determined from a 0.1 wt. % dispersion with dynamic light scattering (DLS) (Litesizer 500, Anton Paar, 175°, λ = 660 nm) with 3 measurements of 60 s and cumulant analysis fit. Fig. S1c shows the hydrodynamic radius $R_h$ found for each temperature. The CS microgels decrease in size from $R_h$ = 236.6 nm at 15 °C to $R_h$ = 151.1 nm at 50 °C, which indicates a 75% volume decrease over the full temperature range. With a sigmoidal fit the VPTT transition was found at T = 32.3 °C, which is in agreement with typical VPTT values for aqueous PNIPAM microgels.[1] In addition, the zeta-potential ζ was determined from electrophoretic measurements. Fig. S1d shows ζ over the full temperature range and indicates that ζ decreases from ζ ~ -10 mV to ζ ~ -30 mV as the temperature is raised from T = 20 °C to T = 40 °C caused by the PNIPAM shell collapse and increase of surface charges.

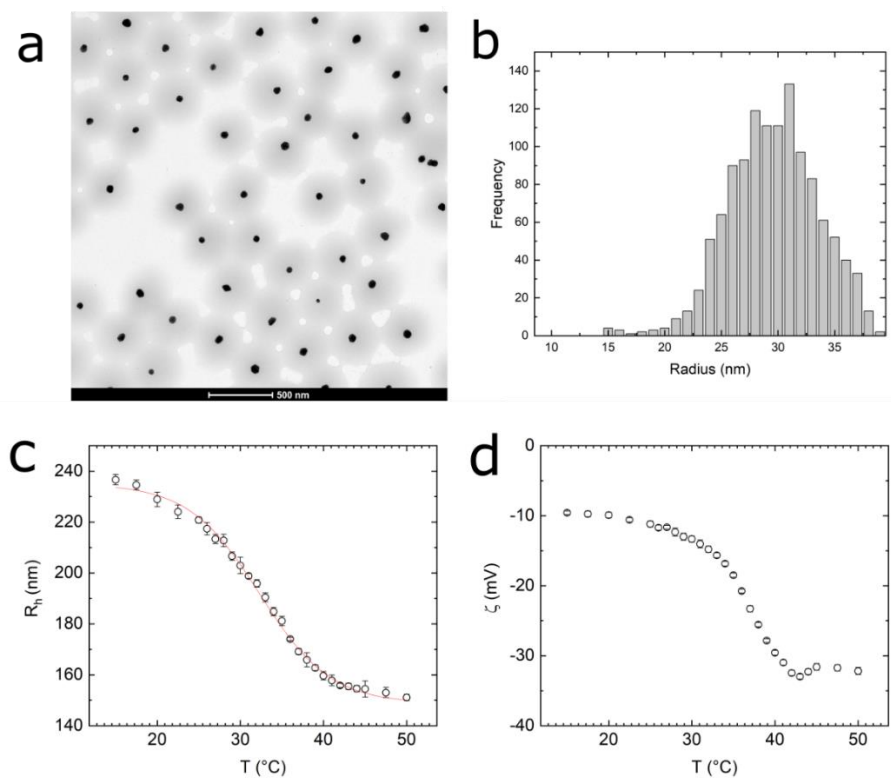

**Figure S1.** (a) Representative TEM images of the core-shell microgels with a gold core and hydrogel shell. (b) Size distribution of $R_{core}$. (c) Hydrodynamic radius $R_h$ with sigmoidal fit (red line) and (d) the zeta-potential of the CS microgels in the range of T = 15 – 50 °C. Error bars represent measured polydispersity index (c) and standard deviation (d).

## S2. Volume fraction determination

Particle dispersions were prepared at 0.5 wt% and 12 wt%. The temperature dependent volume fractions $\phi_{eff}$ (T) were determined from the number density, $N$, based on interparticle distance, $a$, extracted from the Bragg peak position in the SAXS patterns in the full crystallized state of the 12 wt% sample and the CS particle volume. For this we assume an overall FCC lattice and base the particle volume on the hydrodynamic radius, $R_h$(T), as measured with the DLS (Fig. S1c). The effective volume fraction is calculated according to:

$$\phi_{eff}(T) = NV_{CS} = (V_{FCC}/4)V_{CS}(T) = \left((2a/\sqrt{2})^3/4\right)(4/3\,\pi R_h(T)^3), \tag{S1}$$

where $V_{cs}$(T) is the CS particle volume controlled by temperature, and $V_{FCC}$ is the *fcc* unit cell volume. At T = 38 °C we find $a$ = 490 nm and $R_h$ = 165.8 nm, resulting in $\phi_{eff}$ = 0.23. Table S1 shows an overview of the changing $\phi_{eff}$ with temperature based on this calculation. This shows that at T = 20 °C with $R_h$ = 228.9 nm the system has a volume fraction of $\phi_{eff}$ = 0.60. By correcting for wt% we can extract $\phi_{eff}$ for the 0.5 wt% sample resultin in $\phi_{eff}$ = 0.025 at T = 20 °C and $\phi_{eff}$ = 0.010 at T = 38 °C.

## S3. Temperature controlled sample holder

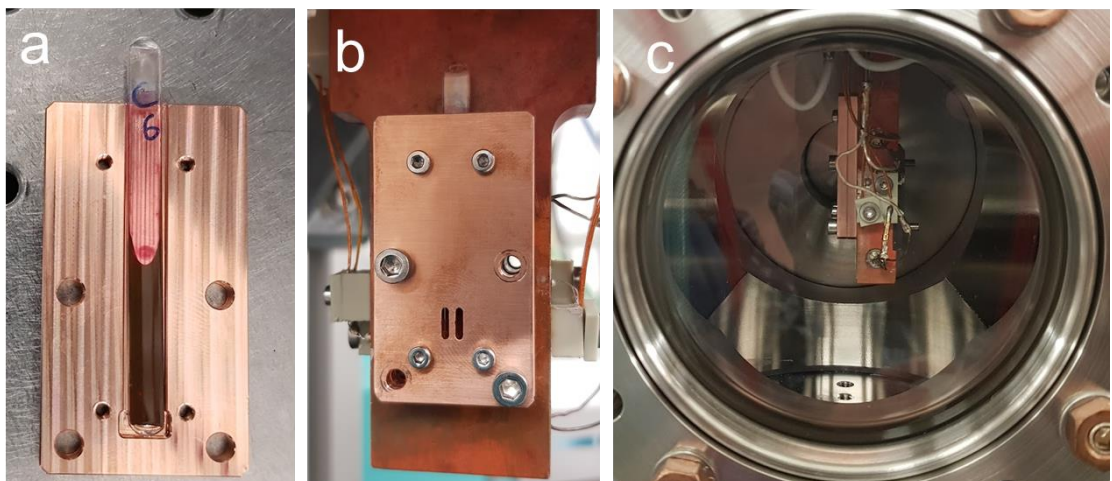

**Figure S2.** Photographs of (**a**) one part of a copper sample holder that had an cavity into which the flame sealed capillary (4 × 0.2 × 50 mm³, internal dimensions, Vitrocom) with CS particle dispersion fitted tightly, (**b**) the full copper sample holder with capillary mounted onto the Peltier element, (**c**) side view of sample holder in vacuum tube.

## S4. Scattering from the dilute sample

We investigated the CS microgels in the dilute state with a concentration of 0.5 wt. % corresponding to a volume fraction $\phi_{eff}$(20 °C) = 0.025. Measurements were performed in the temperature range of 25.0 °C < T < 50.0 °C. Details on the Ultra-Small Angle X-ray Scattering are given in Experimental section of the main text.

The theoretical description of the X-ray scattering is given according to the literature.[2,3] The scattered intensity in kinematical approximation from a dilute sample, where interference between scattering from different particles can be neglected (i.e. the structure factor S(q) ≈ 1) can be described as[2]

$$I(q) = N\Delta\rho^2 V^2 P(q), \tag{S2}$$

where $N$ is the number of illuminated particles in the scattering volume, $\Delta\rho = \rho - \rho_{solvent}$ is the scattering contrast (the electron density difference between the particles and medium (solvent)), $V$ is the volume and $P(q) = |F(q)|^2$ the form factor of a single microgel.

The form factor amplitude for a homogeneous sphere with the radius $R$ is[2]

$$F_1(q, R) = \frac{3[\sin(qR) - qR\cos(qR)]}{(qR)^3}, \tag{S3}$$

The polydispersity of the particles can be introduced by averaging over the particle radius distribution. For the normal size distribution, the probability density is

$$D(R, \langle R \rangle, \sigma_R) = \frac{1}{\sqrt{2\pi\sigma_R^2}} \exp\left[-\frac{(R - \langle R \rangle)^2}{2\sigma_R^2}\right]. \tag{S4}$$

where $\langle R \rangle$ is the mean radius and $\sigma_R$ is the standard deviation. Then, the scattered intensity can be expressed as[3]

$$I(q) = N\Delta\rho^2 \int_0^\infty D(R) V^2(R) |F_1(q, R)|^2 dR. \tag{S5}$$

The form factor amplitude for a core-shell particle can be obtained by a proper weighting of the partial amplitudes from the spherical core and shell as[3]

$$F_2(q, R_{core}, R_{shell}) = \frac{(\Delta\rho_{core} - \Delta\rho_{shell})V(R_{core})F_1(q, R_{core}) + \Delta\rho_{shell}V(R_{shell})F_1(q, R_{shell})}{(\Delta\rho_{core} - \Delta\rho_{shell})V(R_{core}) + \Delta\rho_{shell}V(R_{shell})}, \tag{S6}$$

where $R_{core}$ and $R_{shell}$ are the radii, and $\Delta\rho_{core}$ and $\Delta\rho_{shell}$ are the scattering contrasts of the core and shell, respectively. The particle volume V(R) in this case is equal to the sphere volume $V(R) = \frac{4}{3}\pi R^3$.

By analogy, the polydispersity can be taken into account if one considers the core and shell radii distributions. Then, the resulting scattered intensity can be defined as

$$I(q) = N \int_0^\infty \int_0^\infty D(R_{core}, \langle R_{core} \rangle, \sigma_{R_{core}}) D(R_{shell}, \langle R_{shell} \rangle, \sigma_{R_{shell}}) \times$$
$$\times [(\Delta\rho_{core} - \Delta\rho_{shell})V(R_{core}) + \Delta\rho_{shell}V(R_{shell})]^2 |F_2(q, R_{core}, R_{shell})|^2 dR_{core} dR_{shell} \tag{S7}$$

where $\langle R_{core} \rangle$ and $\langle R_{shell} \rangle$ are the mean radii and $\sigma_{R_{core}}$ and $\sigma_{R_{shell}}$ are the standard deviation of the radii of the core and shell, respectively.

We define an effective form factor for polydisperse core-shell particles as

$$P_{cs}(q) = \frac{1}{\langle (\Delta\rho V)^2 \rangle} \int_0^\infty \int_0^\infty D(R_{core}, \langle R_{core} \rangle, \sigma_{R_{core}}) D(R_{shell}, \langle R_{shell} \rangle, \sigma_{R_{shell}}) \times$$
$$\times [(\Delta\rho_{core} - \Delta\rho_{shell})V(R_{core}) + \Delta\rho_{shell}V(R_{shell})]^2 |F_2(q, R_{core}, R_{shell})|^2 dR_{core} dR_{shell} \tag{S8}$$

where

$$\langle(\Delta\rho V)^2\rangle = \int_0^\infty \int_0^\infty D(R_{core}, \langle R_{core}\rangle, \sigma_{R_{core}}) D(R_{shell}, \langle R_{shell}\rangle, \sigma_{R_{shell}}) \times \quad (S9)$$

$$\times [(\Delta\rho_{core} - \Delta\rho_{shell})V(R_{core}) + \Delta\rho_{shell}V(R_{shell})]^2 dR_{core} dR_{shell}.$$

We fitted the radially-averaged intensity profiles of the diffraction patterns from the dilute samples shown in Fig. S2a,b. We used the polydisperse core-shell form factor given in Eq. S7 multiplied by the scaling parameter $I_0$:

$$I(q) = I_0 P_{cs}(q) \quad (S10)$$

The scattering contrast values were calculated in respect to water with the electron density $\rho_{H_2O}$ = 335 nm$^{-3}$. The gold core electron density was considered to be $\rho_{Au}$ = 4661 nm$^{-3}$. This gives a core scattering contrast of $\Delta\rho_{core}$ = 4326 nm$^{-3}$ which was fixed in the fitting procedure.[4] All other parameters, namely, the scaling parameter $I_0$, the shell scattering contrast $\Delta\rho_{shell}$, the mean core ($\langle R_{core}\rangle$) and shell radii ($\langle R_{shell}\rangle$) and their standard deviations $\sigma_{R_{core}}$ and $\sigma_{R_{shell}}$, respectively, were used as fitting parameters to fit the intensity profiles at different temperatures T. Two examples of the fit at two characteristic temperatures of T = 25.0 °C and T = 50.0 °C are shown in Fig. S2c,d. By that, we extracted these parameters for the dilute sample measured at different temperatures in the range of T = 25.0 – 50.0 °C. The extracted parameters are shown in Fig. S3.

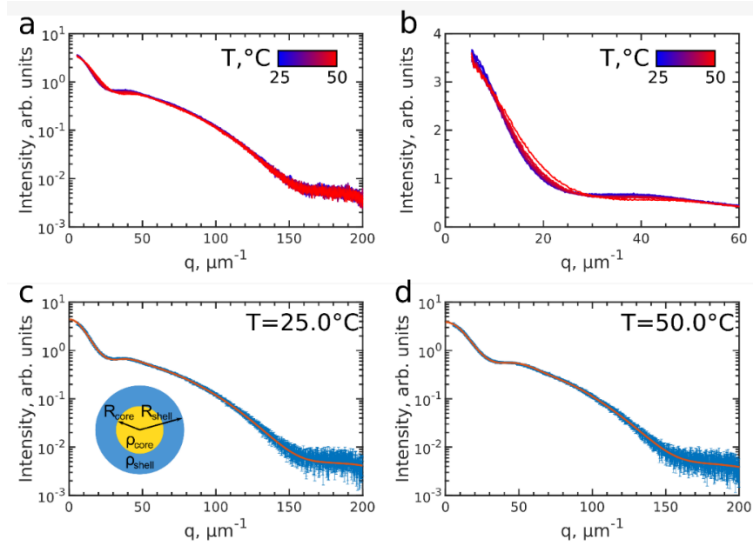

**Figure S3. (a)** Radially averaged intensity from the dilute sample at different temperatures in the range of T = 25.0 – 50.0 °C and **(b)** magnification of low-q area showing the change in the shell scattering. The error bars are omitted for clarity. **(c,d)** Examples of the measured intensities (blue lines) and the best fits with the core-shell model. for the two extreme temperatures: (c) T = 25.0 °C and (d) T = 50.0 °C. Inset in (c) shows the model parameters used for fitting.

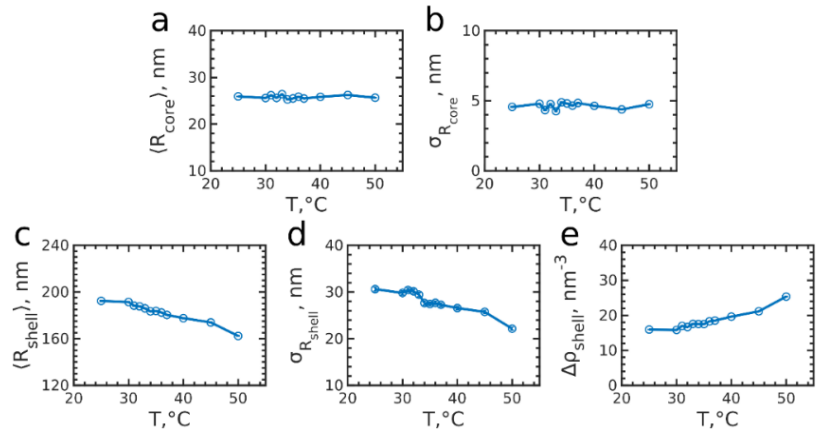

**Figure S4.** Evolution of the extracted parameters for the core-shell microgels with the temperature. **(a)** The mean gold core radius $\langle R_{core}\rangle$ and **(b)** its standard deviation $\sigma_{R_{core}}$, **(c)** mean shell radius $\langle R_{shell}\rangle$, **(d)** its standard deviation $\sigma_{R_{shell}}$, and **(e)** shell scattering contrast $\Delta\rho_{shell}$.

## S5. Fluid structure factor

We now turn to the densely packed microgel sample with the concentration of 12 wt. % (an effective volume fraction $\phi_{eff}$(20 °C) = 0.60). Due to the volume phase transition behaviour of the microgels, this sample will possess a significantly reduced effective volume fraction at high temperatures. Here, the sample reveals a typical fluid-like intensity profile, as shown in the main text in Fig. 3d and Fig. 6d. One of the analytically calculated structure factor models for a fluid is a hard-sphere model in the Percus-Yevick approximation.[3] In this model, the particles interact with the hard-sphere radius $R_{PY}$ and have a hard-sphere volume fraction $\phi_{PY}$. Then, the structure factor can be expressed as

$$S(q) = \frac{1}{1 + 24\phi_{PY}\frac{G(2qR_{PY})}{2qR_{PY}}}, \tag{S11}$$

where

$$G(x) = \alpha \frac{\sin x - x \cos x}{x^2} + \beta \frac{2x \sin x + (2 - x^2) \cos x - 2}{x^3} + \gamma \frac{-x^4 \cos x + 4[(3x^2 - 6)\cos x + (x^3 - 6x)\sin x + 6]}{x^5}, \tag{S12}$$

and

$$\alpha = \frac{(1 + 2\phi_{PY})^2}{(1 - \phi_{PY})^4},$$

$$\beta = \frac{-6\phi_{PY}(1 + \phi_{PY}/2)^2}{(1 - \phi_{PY})^4}, \tag{S13}$$

$$\gamma = \alpha\phi_{PY}/2.$$

The experimentally obtained 1D intensity profiles were fitted by this model using the hard sphere radius $R_{PY}$ and the volume fraction $\phi_{PY}$ as the fitting parameters. The results of the fit are shown in the main text in Fig. 3d and Fig. 6d.

## S6. Bragg peak fitting

An example of the diffraction pattern at T = 35 °C is shown in Fig. S4a. The pattern was interpolated into polar $(q,\varphi)$-coordinates and divided by the form factor of the CS nanoparticles $P_{cs}(q)$ extracted from the diluted sample as described in Section S2. At each $q$-value, the intensity between the Bragg peaks $S_{iso}(q)$ was calculated and additionally subtracted from the pattern to keep only the anisotropic Bragg peaks. The resulting pattern in polar coordinates is shown in Fig. S4b.

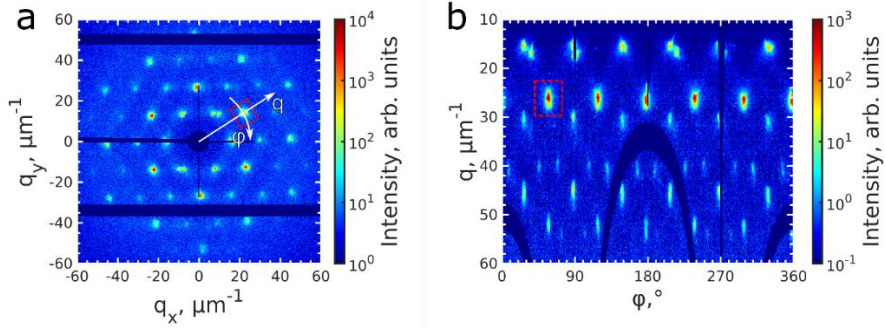

**Figure S5. (a)** Example of the diffraction patterns at T = 35 °C, when the sample is crystallized. **(b)** The same pattern interpolated into polar coordinates after division by the form factor and subtraction of the isotropic part. The highlighted peak is shown enlarged in Fig. S5a.

The peak highlighted in Fig. S4 is shown separately in Fig. S5a. It was fitted by the 2D Gaussian function:

$$G(q,\varphi) = \frac{I}{2\pi\sigma_q\sigma_\varphi}\exp\left[-\frac{(q-q_0)^2}{2\sigma_q^2} - \frac{(\varphi-\varphi_0)^2}{2\sigma_\varphi^2}\right], \qquad (S8)$$

where $I$ is the integrated intensity, $q_0$ and $\varphi_0$ are the peak positions in radial and azimuthal directions, and $\sigma_q$ and $\sigma_\varphi$ are the peak width in radial and azimuthal directions. All these parameters were fitted by the least squares method to the experimental intensities. An example of the fit is shown in Fig. S5b. The radial FWHM was calculated as $w_q = 2\sqrt{2\ln 2}\sigma_q$, the azimuthal one as $w_\varphi = 2\sqrt{2\ln 2}\sigma_\varphi q_0$. The parameters were averaged over all peaks of the corresponding Bragg peak family. The errors were estimated as the standard deviation within a Bragg peak family.

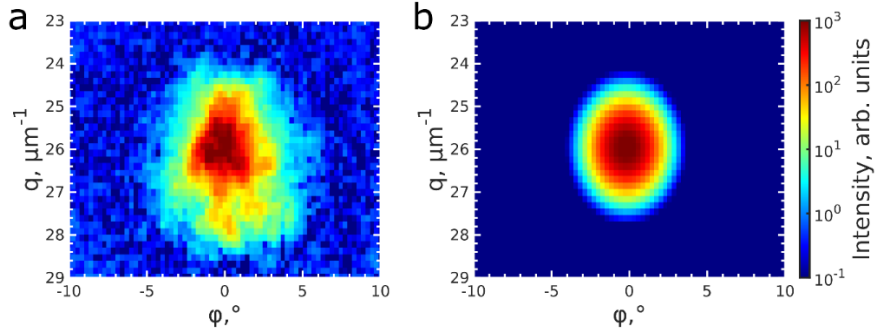

**Figure S6. (a)** Magnified area in the vicinity of one of the Bragg peaks from the $2\bar{1}\bar{1}0$ family. The peak is highlighted in Fig. S4. **(b)** The result of the peak fitting by a 2D Gaussian function.

## S7. Bragg peak fitting while heating

During heating we observed splitting of each Bragg peak into three subpeaks as shown in the main text, Fig. 7e-h. The subpeaks were fitted by a linear combination of three Gaussian functions (eq. S8). Three sets of the parameters were optimized simultaneously. Evolution of the extracted parameters for each subpeak is shown in Fig. S6 – S8.

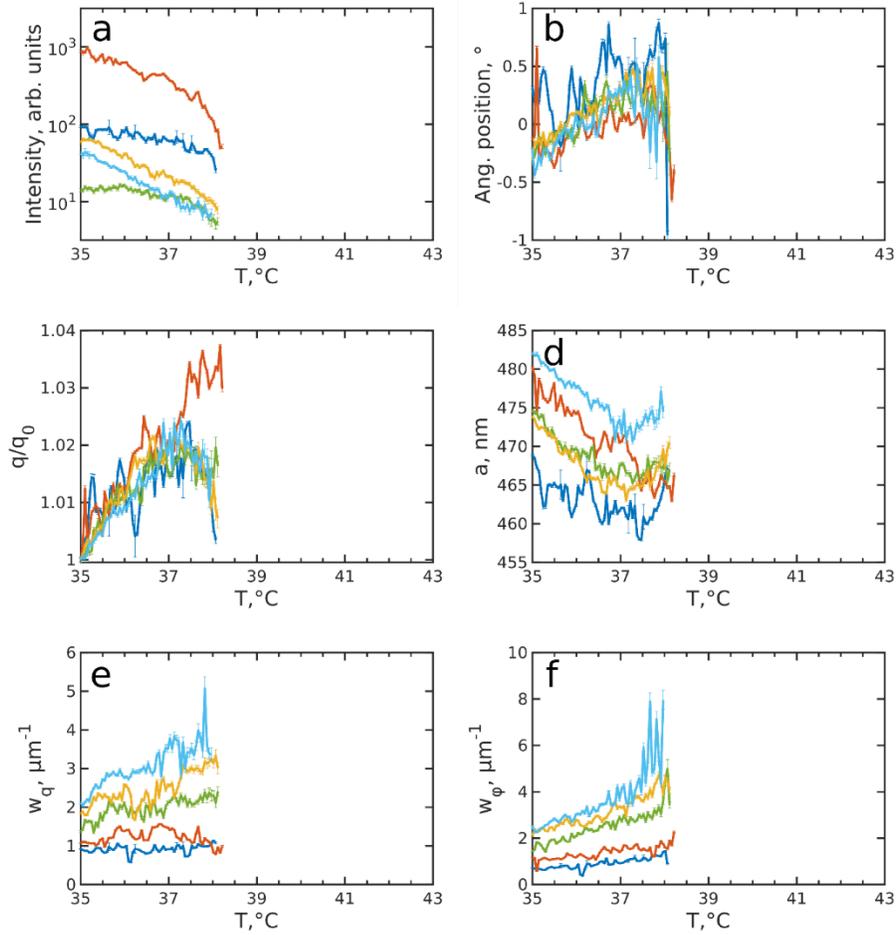

**Figure S7.** Evolution of the parameters of the "blue" subpeaks of different orders: **(a)** the intensity, **(b)** the angular position, **(c)** the momentum transfer values with respect to the initial values, **(d)** the calculated unit cell parameters, **(e)** the FWHM in radial direction and **(f)** the FWHM in azimuthal direction.

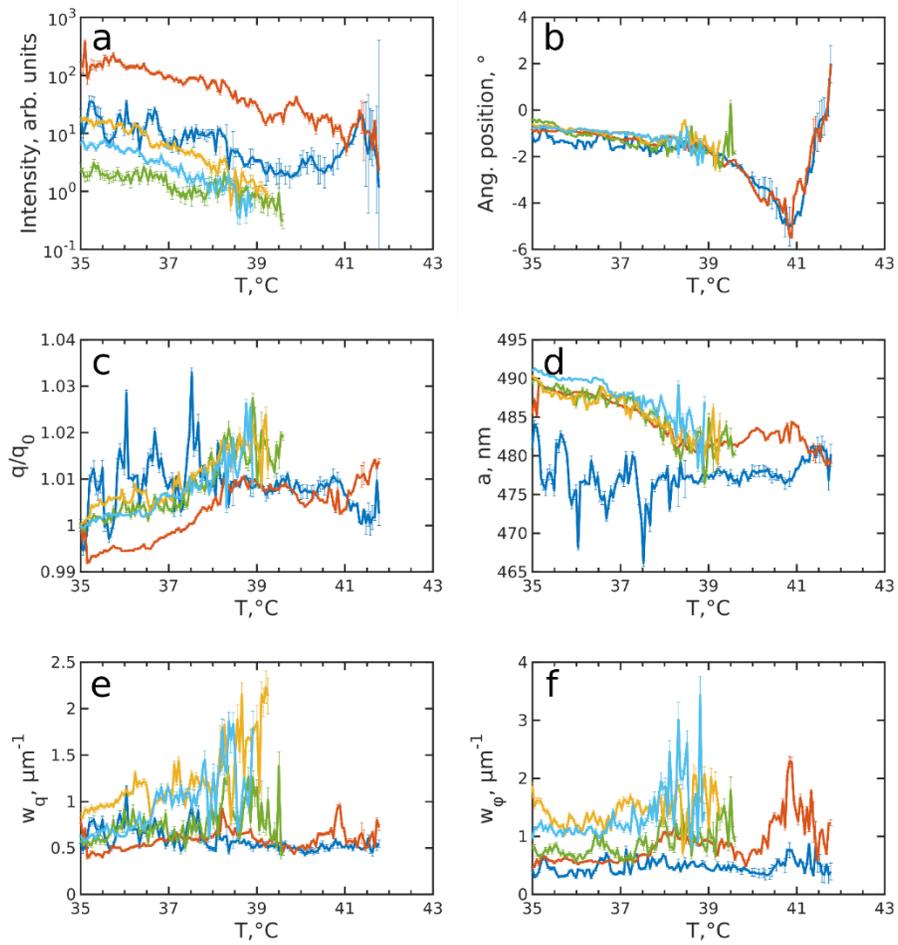

**Figure S8.** Evolution of the parameters of the "red" subpeaks of different orders: **(a)** the intensity, **(b)** the angular position, **(c)** the momentum transfer values with respect to the initial values, **(d)** the calculated unit cell parameters, **(e)** the FWHM in radial direction and **(f)** the FWHM in azimuthal direction.

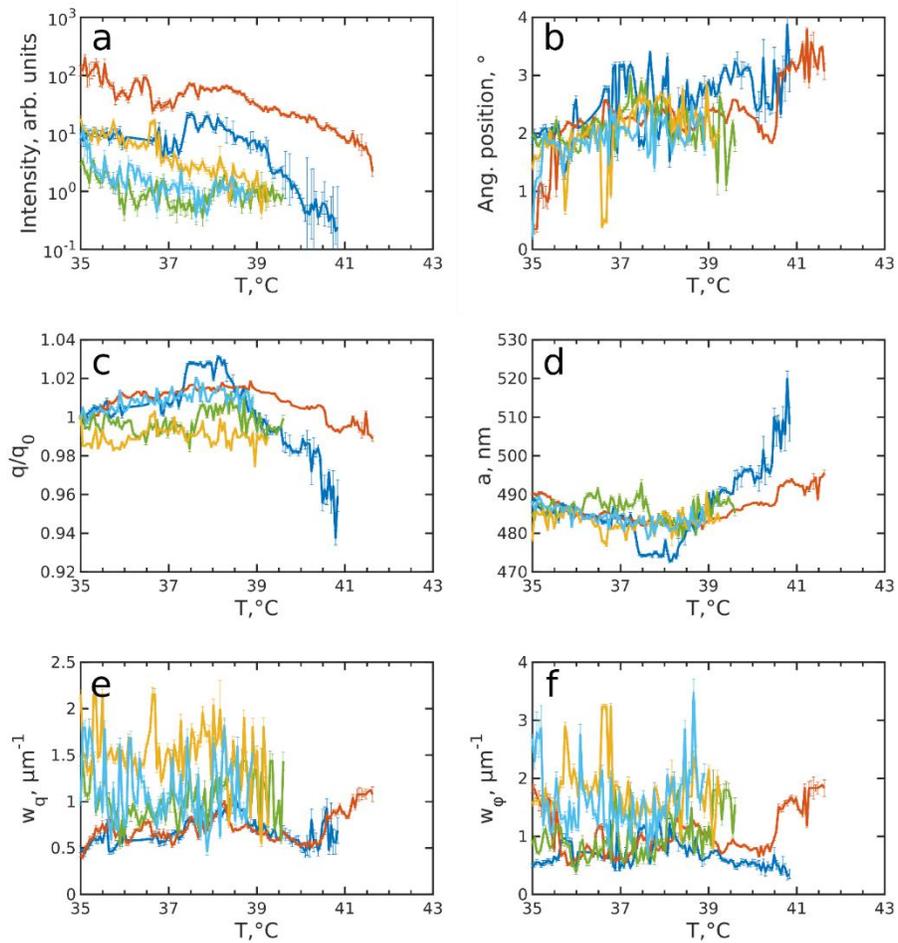

**Figure S9.** Evolution of the parameters of the "green" subpeaks of different orders: **(a)** the intensity, **(b)** the angular position, **(c)** the momentum transfer values with respect to the initial values, **(d)** the calculated unit cell parameters, **(e)** the FWHM in radial direction and **(f)** the FWHM in azimuthal direction.

Besides the lattice spacing analysis, we performed Williamson-Hall analysis. We did find the results were not reliable because the FWHMs fits contained large errors. However, the resulting distortion values we did obtain for each crystallite are shown in Fig. S9 and again indicate significant difference in the crystallite behaviour. The "blue" crystallite initial distortion values of $g_q$ = 4 ± 1% and $g_\varphi$ = 5 ± 1% in radial and angular directions, respectively, are very similar to the average values at the end of cooling. During further heating, the distortions monotonically increase and reach the values of $g_q$ = 7 ± 2% and $g_\varphi$ = 10 ± 2% at T ≈ 38.2 °C, when the peaks disappear, indicating distortions in the crystallite structure grow. For the "red" and "green" crystallites, the values are initially lower at the level of $g$ ≈ 3 ± 1% in both directions and stay almost constant in the whole temperature range until the crystallites fully melt. The latter shows that these crystals do not distort during melting.

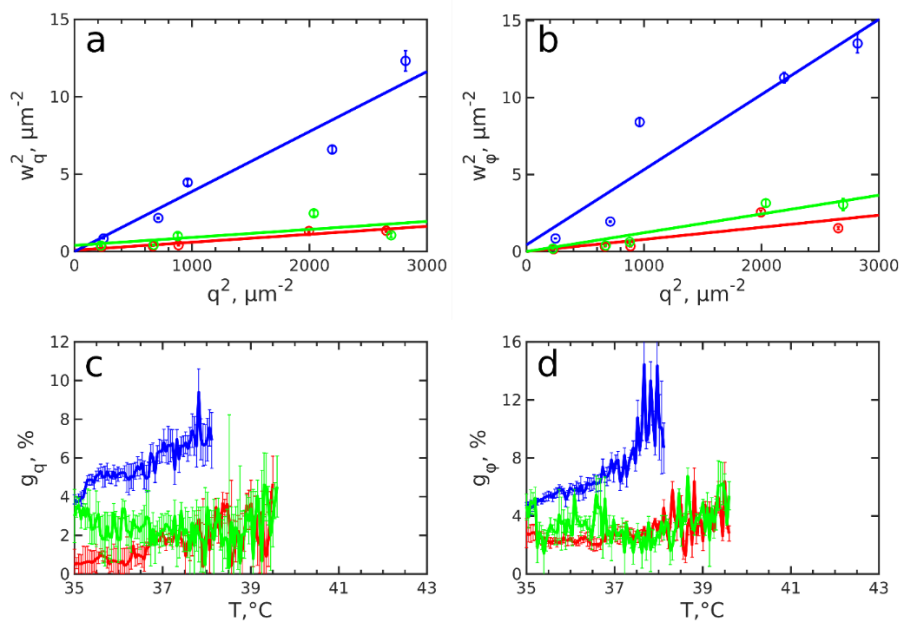

**Figure S10. (a,b)** Examples of the Williamson-Hall plots for the FWHMs of the subpeaks in (a) radial and (b) azimuthal directions at T = 37.0 °C. Points are experimental values and straight lines are the best fits. **(c,d)** Evolution of the extracted (c) radial and (d) angular lattice distortions during heating.

TABLE OF CONTENT

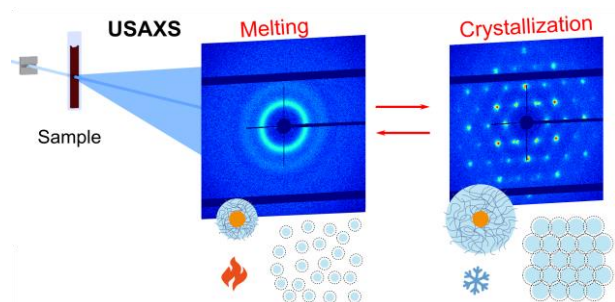

By combining thermoresponsive core-shell gold-PNIPAM microgels with USAXS, the crystallization and melting of soft colloidal crystals is investigated in detail with Bragg peak analysis and reveals complex crystal growth and melting in different stages.